\documentclass[smallextended]{svjour3}
\usepackage{graphicx, amsmath, amssymb}
\usepackage[caption=false]{subfig}

\newcommand{\braket}[2]{{\left\langle #1 \middle| #2 \right\rangle}}

\newcommand{\ket}[1]{{\left| #1 \right\rangle}}
\newcommand{\ketbra}[2]{{\left| #1 \middle\rangle \middle \langle #2 \right|}}

\journalname{Quantum Inf Process}

\begin{document}

\title{Faster Search by Lackadaisical Quantum Walk}

\author{Thomas G.~Wong}

\authorrunning{T.~G.~Wong}

\institute{T.~G.~Wong \at
	   Department of Computer Science, University of Texas at Austin, 2317 Speedway, Austin, Texas 78712, USA \\
	   Currently at Department of Physics, Creighton University, 2500 California Plaza, Omaha, NE 68178, USA \\
	   \email{thomaswong@creighton.edu}
}

\date{Received: date / Accepted: date}

\maketitle

\begin{abstract}
	In the typical model, a discrete-time coined quantum walk searching the 2D grid for a marked vertex achieves a success probability of $O(1/\log N)$ in $O(\sqrt{N \log N})$ steps, which with amplitude amplification yields an overall runtime of $O(\sqrt{N} \log N)$. We show that making the quantum walk lackadaisical or lazy by adding a self-loop of weight $4/N$ to each vertex speeds up the search, causing the success probability to reach a constant near $1$ in $O(\sqrt{N \log N})$ steps, thus yielding an $O(\sqrt{\log N})$ improvement over the typical, loopless algorithm. This improved runtime matches the best known quantum algorithms for this search problem. Our results are based on numerical simulations since the algorithm is not an instance of the abstract search algorithm.
\end{abstract}


\section{Introduction}

Grover's quantum search algorithm \cite{Grover1996} famously searches an unordered database of $N$ items in $O(\sqrt{N})$ time, which is a quadratic improvement over the $O(N)$ steps that a classical computer would need to take. Fifteen years ago, however, Benioff \cite{Benioff2002} noted that a quantum computer may lose this speedup when searching a spatial region since it takes time for a ``quantum robot'' to traverse the database.

Since then, much work has explored how quickly quantum computers can search spatial regions, beginning with Aaronson and Ambainis \cite{AA2005}, who gave a recursive algorithm that searches the two-dimensional (2D) grid, or periodic square lattice or torus, of $N$ vertices in $O(\sqrt{N} \log^2 N)$ steps while achieving the full Grover speedup of $O(\sqrt{N})$ in higher-dimensional lattices. Another approach is using a continuous-time quantum walk, which Childs and Goldstone \cite{CG2004a} showed searches no better than classical in 2D and 3D, in $O(\sqrt{N} \log^{3/2} N)$ in 4D, and in $O(\sqrt{N})$ in 5D or greater (see Table 1 of \cite{Wong7} for a detailed summary). Ambainis \textit{et al.}~\cite{AKR2005} then showed that a discrete-time coined quantum walk outperforms both of the previous algorithms, searching the 2D grid in $O(\sqrt{N} \log N)$ steps and higher-dimensional lattices in $O(\sqrt{N})$. Specifically, for the 2D grid, the probability at the marked vertex reaches $O(1/\log N)$ after $O(\sqrt{N \log N})$ steps. Using amplitude amplification \cite{BHMT2000}, this implies a search algorithm with an overall runtime of $O(\sqrt{N} \log N)$. Childs and Goldstone later matched these runtimes with a continuous-time quantum walk governed by the Dirac equation \cite{CG2004b}.

\begin{figure}
\begin{center}
	\includegraphics{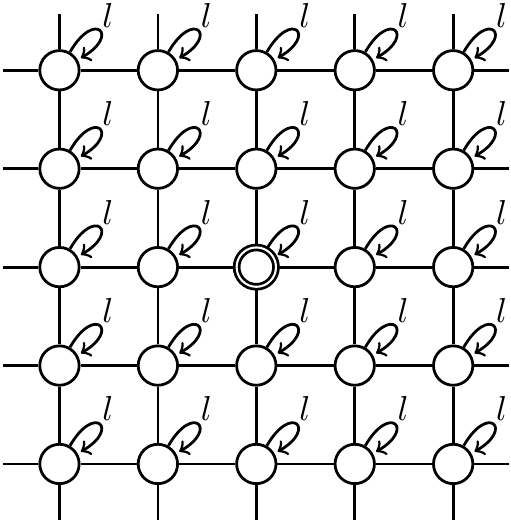}
	\caption{\label{fig:grid} A 2D grid of $N = 5 \times 5$ vertices with a self-loop of weight $l$ at each vertex. The boundaries are periodic. A marked vertex is indicated by a double circle.}
\end{center}
\end{figure}

In this paper, we investigate a simple modification to the discrete-time quantum walk search algorithm of Ambainis \textit{et al.}~\cite{AKR2005} on the 2D grid, where we make the walker lazy or lackadaisical \cite{Wong10,Wong27}. We do this by giving each vertex of the 2D grid a self-loop of weight $l$, as illustrated in Fig.~\ref{fig:grid}, so the walker has some probability of staying put. We perform a discrete-time coined quantum walk on this weighted graph while querying a Grover-type oracle that flips the sign of the amplitude at the marked vertex. Numerically, when $l = 4/N$, the probability at the marked vertex reaches a constant after $O(\sqrt{N \log N})$ steps, which is an $O(\sqrt{\log N})$ improvement over the loopless algorithm's $O(\sqrt{N} \log N)$ \cite{AKR2005}.

This indicates that a lackadaisical quantum walk is capable of giving speedups in the runtime scaling, not just speedups in the constant factor \cite{Wong10,Wong27}, and it suggests that self-loops might be useful algorithmic tools in other quantum walk algorithms as well, not just those for searching. Note there are other approaches to achieving this improved runtime of $O(\sqrt{N \log N})$, such as controlling the walk with an ancilla qubit \cite{Tulsi2008}, searching a neighborhood \cite{Ambainis2013}, quantizing a semi-absorbing random walk \cite{Krovi2016}, and using staggered quantum walks with Hamiltonians \cite{Portugal2017}, but the simplicity of adding self-loops to a graph makes lackadaisical quantum walks an attractive alternative.

In the next section, we describe the lackadaisical algorithm in detail and give numerical results. Then we describe why this algorithm is unamenable to the usual analytical method of the abstract search algorithm \cite{AKR2005}. Finally, we end with concluding remarks.


\section{Search by Lackadaisical Quantum Walk}

In a quantum walk, the $N$ vertices of a graph form a computational basis $\{ \ket{1}, \ket{2}, \dots, \ket{N} \}$ for $\mathbb{C}^N$. A discrete-time quantum walk, however, also includes an additional coin degree of freedom encoding the directions in which a particle can hop \cite{Meyer1996a,Meyer1996b}. For the 2D grid with self-loops, as in Fig.~\ref{fig:grid}, the directions are $\{ \ket{\uparrow}, \ket{\downarrow}, \ket{\leftarrow}, \ket{\rightarrow}, \ket{\circlearrowleft} \}$, and these form a computational basis for $\mathbb{C}^5$. So the Hilbert space of the quantum walk is $\mathbb{C}^N \otimes \mathbb{C}^5$. For example, the state $\ket{2,\rightarrow}$ denotes a particle at vertex 2 pointing to the right. The pure quantum walk (without searching) evolves by repeated applications of
\[ U = S \cdot (I_N \otimes C), \]
where
\[ C = 2 \ketbra{s_c}{s_c} - I_5 \]
with
\[ \ket{s_c} = \frac{1}{\sqrt{4+l}} \left( \ket{\uparrow} + \ket{\downarrow} + \ket{\leftarrow} + \ket{\rightarrow} + \sqrt{l} \ket{\circlearrowleft} \right) \]
is the Grover diffusion coin for a weighted graph \cite{Wong27}, and $S$ is the flip-flop shift that causes a particle to hop and turn around \cite{AKR2005}. The system begins in 
\begin{equation}
	\label{eq:start}
	\ket{\psi_0} = \frac{1}{\sqrt{N}} \sum_{v = 1}^N \ket{v} \otimes \ket{s_c},
\end{equation}
which is a uniform distribution over the vertices (but not necessarily the directions). Note this is the unique eigenvector of the quantum walk operator $U$ with eigenvalue $1$, and it can be prepared in $O(\sqrt{N})$ steps without knowing the marked vertex.

Now to search, we include a query to an oracle with each step of the quantum walk. Two choices for the oracle are common. The first is the ``Grover oracle,'' where the algorithm repeatedly applies
\begin{equation}
	\label{eq:U'}
	U' = U \cdot (Q \otimes I_5).
\end{equation}
Here, $Q = I_N - 2\ketbra{w}{w}$, where $\ket{w}$ denotes the marked vertex. This oracle flips the sign at the marked vertex, irrespective of the coin state. $U'$ is equivalent to applying $C$ to unmarked vertices and $-C$ to the marked vertex, followed by the shift \cite{AKR2005}. Another choice is the ``SKW oracle'' \cite{SKW2003}, where the algorithm applies
\begin{equation}
	\label{eq:U''}
	U'' = U \cdot (I_{5N} - 2 \ketbra{w,s_c}{w,s_c}).
\end{equation}
Now the oracle term only exactly flips the sign at the marked vertex if the coin state is $\ket{s_c}$. $U''$ is equivalent to applying $C$ to unmarked vertices and $-I_5$ to the marked vertex, followed by the shift \cite{AKR2005}. It is also equivalent to search by Szegedy's quantum walk \cite{Szegedy2004,Magniez2012,Wong26}.

\begin{figure}
\begin{center}
	\subfloat[]{
		\includegraphics{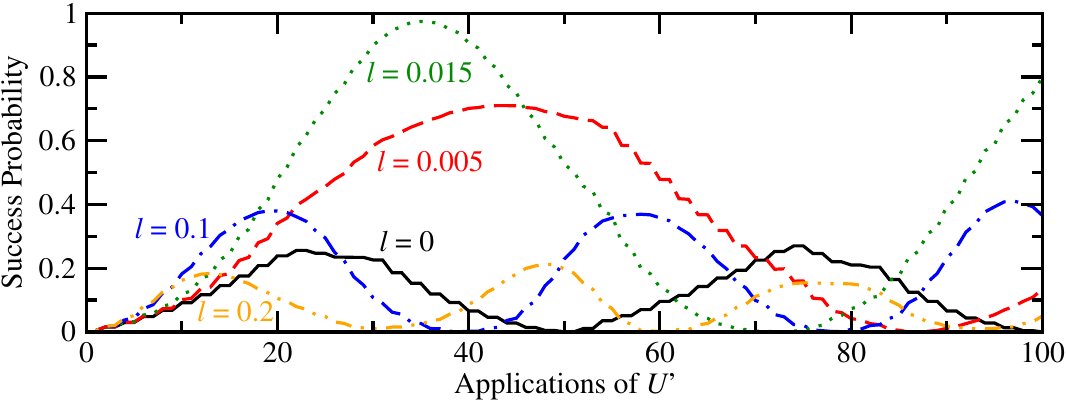}
		\label{fig:prob_time_Grover_L16} 
	}

	\subfloat[]{
		\includegraphics{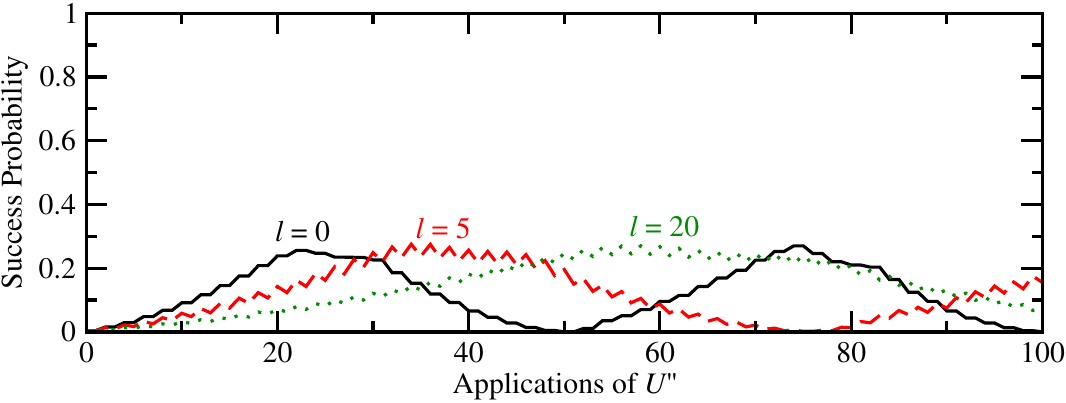}
		\label{fig:prob_time_SKW_L16} 
	}
	\caption{\label{fig:prob_time_L16} Success probability as a function of time for search on the 2D grid of $N = 16 \times 16 = 256$ vertices with a self-loop of weight $l$ at each vertex. (a) With the Grover oracle, the solid black curve is $l = 0$, the dashed red curve is $l = 0.005$, the dotted green curve is $l = 0.015$, the dot-dashed blue curve is $l = 0.1$, and the dot-dot-dashed orange curve is $l = 0.2$. (b) With the SKW oracle, the solid black curve is $l = 0$, the red dashed curve is $l = 5$, and the dotted green curve is $l = 20$.}
\end{center}
\end{figure}

In Fig.~\ref{fig:prob_time_L16}, we simulate the search algorithm with each oracle for a grid of size $N = 16 \times 16 = 256$ with various values of $l$. Figure~\ref{fig:prob_time_Grover_L16} uses the Grover oracle \eqref{eq:U'}, and when $l = 0$, this reproduces the loopless quantum walk of \cite{AKR2005}, where the success probability reaches a value of $O(1/\log N)$ at time $O(\sqrt{N \log N})$. As we increase $l$, the success probability improves, almost reaching $1$ when $l = 0.015$. This probability is almost entirely contained in the marked vertex's self-loop, so the algorithm ``stores'' success probability in the self-loop. As $l$ increases further, however, the success probability drops, so there is an optimal amount of ``laziness'' for which the success probability is maximized.

Figure~\ref{fig:prob_time_SKW_L16} uses the SKW oracle \eqref{eq:U''}. Again with $l = 0$, this reproduces the loopless algorithm. Now as we increase $l$, the algorithm simply slows down. This lack of improvement is unsurprising, however, since $U''$ is equivalent to applying $-I_5$ to the marked vertex. Then the amplitude in the marked vertex's self-loop simply alternates signs with each step, so the algorithm cannot store amplitude in the marked vertex's self-loop.

\begin{figure}
\begin{center}
	\includegraphics{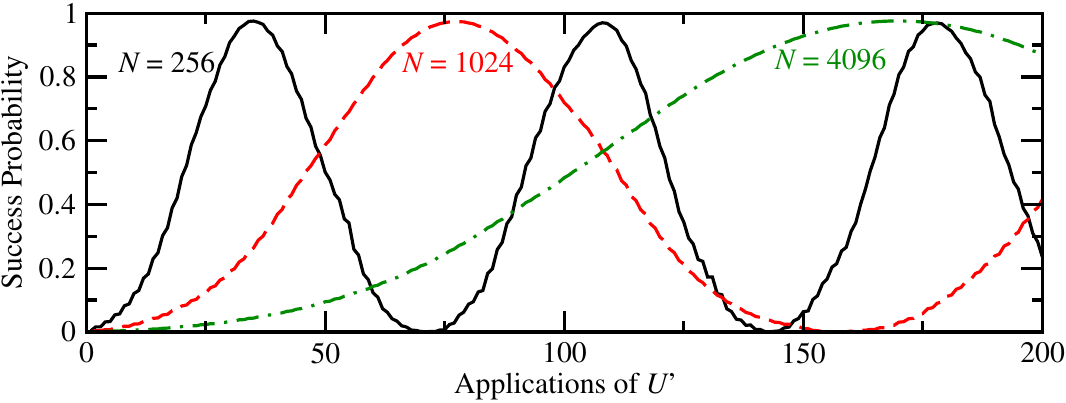}
	\caption{\label{fig:prob_time_Grover_critical} Success probability as a function time for search with the Grover coin on the 2D grid of $N$ vertices with a self-loop of weight $l = 4/N$ at each vertex. The solid black curve is $N = 16 \times 16 = 256$, the dashed red curve is $N = 32 \times 32 = 1024$, and the dotted green curve is $N = 64 \times 64 = 4096$.}
\end{center}
\end{figure}

So let us focus on the quantum walk with the Grover coin \eqref{eq:U'}. Note when $N = 16 \times 16 = 256$, the optimal self-loop weight of roughly $0.015$ is approximately $4/N$. So in Fig.~\ref{fig:prob_time_Grover_critical}, we simulate search on grids of various sizes with $l = 4/N$. We see that in all cases, the success probability is near 1. Specifically, when $N = 16 \times 16 = 256$, the success probability reaches its first peak of $0.975506$ after $35$ steps; when $N = 32 \times 32 = 1024$, the its first peak is $0.973669$ after $77$ steps; and when $N = 64 \times 64 = 4096$, the success probability reaches its first peak of $0.975548$ at $170$ steps.

\begin{figure}
\begin{center}
	\subfloat[]{
		\includegraphics{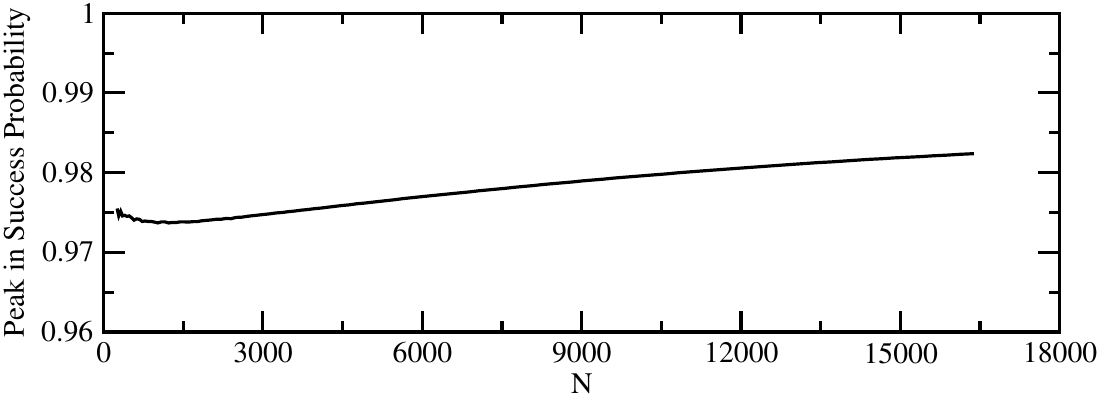}
		\label{fig:firstmax_N_prob} 
	}

	\subfloat[]{
		\includegraphics{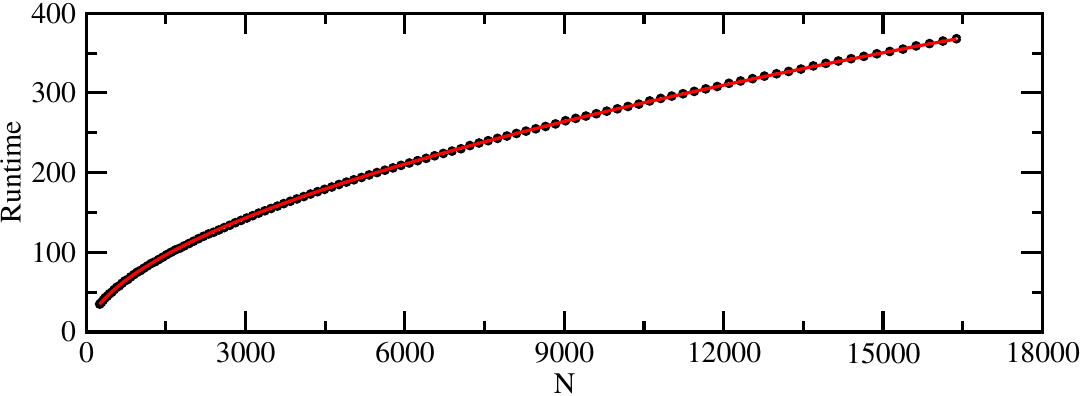}
		\label{fig:firstmax_N_runtime}
	}
	\caption{\label{fig:firstmax} Search with the Grover coin on the 2D grid of $N$ vertices, with $\sqrt{N} = \{ 16, 17, \dots, 128 \}$, by lackadaisical quantum walk with $l = 4/N$. (a) The value of the first peak in success probability. (b) The time to the first peak of success probability. The black dots are the values from numerically simulating the walk. The red curve is the fit $t_* = 0.922466 \sqrt{N \log N}$.}
\end{center}
\end{figure}

We can similarly find the success probability's first peak and the time it occurs for grids of length $\sqrt{N} = \{ 16, 17, \dots, 128 \}$. The complete data and the Python script used to simulate the quantum walk are available in this paper's arXiv source. The peak success probability is shown in Fig.~\ref{fig:firstmax_N_prob}, and it converges to a constant. The time to this peak, or runtime $t_*$, is shown in Fig.~\ref{fig:firstmax_N_runtime}, and the fit
\[ t_* = 0.922466 \sqrt{N \log N} \]
has a correlation coefficient of 0.999993, indicating that the fit is sound. Thus, the numerical simulations suggest that the overall runtime is $O(\sqrt{N \log N})$, which is an $O(\sqrt{\log N})$ improvement over the typical, loopless algorithm's $O(\sqrt{N} \log N)$ in \cite{AKR2005}.


\section{Comparison with Abstract Search Algorithm}

The behavior of the typical, loopless algorithm was analytically determined using a framework known as the abstract search algorithm \cite{AKR2005}. This framework has also been used to analyze several other quantum walk search algorithms \cite{Tulsi2008,Abal2010,Marquezino2013}. Here, we review the abstract search algorithm \cite{AKR2005} and its generalization by Tulsi \cite{Tulsi2008} and show that our lackadaisical quantum walk with the Grover coin \eqref{eq:U'} does not fit into its framework.

The abstract search algorithm repeatedly applies two operators
\begin{equation}
	\label{eq:abstract}
	U_2 U_1
\end{equation}
with the following properties on two states $\ket{\psi_{\rm start}}$ and $\ket{\psi_{\rm good}}$:
\begin{itemize}
	\item	$U_1$ flips the sign of $\ket{\psi_{\rm good}}$ and does nothing to states orthogonal to it. That is, $U_1 = I - 2 \ketbra{\psi_{\rm good}}{\psi_{\rm good}}$.
	\item	$\ket{\psi_{\rm start}}$ has real amplitudes and is the only 1-eigenvector of $U_2$. That is, $U_2 \ket{\psi_{\rm start}} = \ket{\psi_{\rm start}}$.
	\item	$U_2$ is real (and unitary).
\end{itemize}
Then the behavior of the algorithm can be determined by expanding $\ket{\psi_{\rm good}}$ in the eigenbasis of $U_2$. First, $U_2$ has one eigenvalue $1$, and we denote the corresponding eigenvector $\ket{\Phi_0}$. Second, $U_2$ may have zero or more eigenvalues $-1$, and we denote the corresponding eigenvectors $\ket{\Phi_k}$. Finally, since $U_2$ is real and unitary, its remaining eigenvalues come in pairs of complex conjugate numbers $e^{\pm i \theta_j}$, and we denote the corresponding eigenvectors $\ket{\Phi_j^\pm}$, where $\ket{\Phi_j^+} = \ket{\Phi_j^-}^*$. Then we can write the good state in this complete orthonormal basis:
\[ \ket{\psi_{\rm good}} = a_0 \ket{\Phi_0} + \sum_k a_k \ket{\Phi_k} + \sum_j \left( a_j^+ \ket{\Phi_j^+} + a_j^- \ket{\Phi_j^-} \right). \]
Since the good state is a real vector, $a_j^+ = (a_j^-)^*$, and we can choose the phase of $\ket{\Phi_j^\pm}$ so that $a_j^+ = a_j^- = a_j$. Then we have
\[ \ket{\psi_{\rm good}} = a_0 \ket{\Phi_0} + \sum_k a_k \ket{\Phi_k} + \sum_j a_j \left( \ket{\Phi_j^+} + \ket{\Phi_j^-} \right). \]
Now we can use these coefficients (i.e., $a_0$, the $a_k$'s, and the $a_j$'s) to determine the behavior of the abstract search algorithm. Typically, the starting and good states are close to two other states, $\ket{w_{\rm start}}$ and $\ket{w_{\rm good}}$, with respective overlaps
\begin{eqnarray*}
	\left| \braket{\psi_{\rm start}}{w_{\rm start}} \right| \ge 1 - \Theta \left( \frac{\alpha^4}{a_0^2} \sum_j \frac{a_j^2}{(1 - \cos \theta_j)^2} \right) - \Theta \left( \frac{\alpha^4}{a_0^2} \sum_k a_k^2 \right), \\
	\left| \langle \psi_{\rm good} | w_{\rm good} \rangle \right| = \Theta \left( \min \left( \frac{1}{\sqrt{\sum_j a_j^2 \cot^2 \frac{\theta_j}{4}}}, 1 \right) \right),
\end{eqnarray*}
where
\[ \alpha = \Theta \left( \frac{a_0}{\sqrt{\sum_j \frac{a_j^2}{1-\cos\theta_j} + \frac{1}{4} \sum_k a_k^2}} \right). \]
These two states are important because $\pi/2\alpha$ applications of $U_2 U_1$ drives the system from $\ket{w_{\rm start}}$ to $\ket{w_{\rm good}}$. That is, $\ket{w_{\rm good}} = (U_2 U_1)^{\pi/2\alpha} \ket{w_{\rm start}}$. This yields the runtime of the algorithm, and the previous overlaps with $\ket{\psi_{\rm start}}$ and $\ket{\psi_{\rm good}}$ give the success probability.

Quantum walk search algorithms are framed as abstract search algorithms by first identifying the quantum walk operator $U = S(I_N \otimes C)$ with $U_2$. $U$ is real and unitary, and the starting state of the quantum walk $\ket{\psi_0}$ corresponds to $\ket{\psi_{\rm start}}$ since it is the unique 1-eigenvector of $U$. Now comparing \eqref{eq:U'} and \eqref{eq:abstract}, we would also like to identify $(Q \otimes I_5)$ with $U_1$. Unfortunately, $\ket{\psi_{\rm good}}$ cannot exist with this identification. For example, if the marked vertex is $\ket{w}$, then $(Q \otimes I_5)$ flips the signs of both $\ket{w,\uparrow}$ and $\ket{w,\circlearrowleft}$. These two states are orthogonal, so $U_1$ should only flip the sign of one state and do nothing to the other. Therefore, our algorithm \eqref{eq:U'} is not amenable to the analysis of the abstract search algorithm.

We note that search with the SKW oracle \eqref{eq:U''} is an instance of the abstract search algorithm since $I_{5N} - 2 \ketbra{w,s_c}{w,s_c}$ can be identified with $U_1$. Then $\ket{w,s_c}$ is $\ket{\psi_{\rm good}}$, and the oracle acts as the identity on all states orthogonal to this. Analyzing this algorithm is uninteresting, however, since the self-loops only slow down the walk, as we saw in Fig.~\ref{fig:prob_time_SKW_L16}.


\section{Conclusion}

We have shown that the lackadaisical quantum walk, where a self-loop of weight $l$ is added to each vertex of the graph, can achieve a scaling improvement when searching the 2D grid with the Grover oracle for a unique marked vertex when $l = 4/N$. At the marked vertex, the algorithm stores amplitude in the self-loop, allowing it to rise to a constant near $1$ compared with the loopless algorithm's $O(1/\log N)$. This eliminates the need for amplitude amplification, so the resulting algorithm has a runtime of $O(\sqrt{N \log N})$, which is an $O(\sqrt{\log N})$ improvement over the typical, loopless algorithm. This algorithm cannot be formulated in the framework of the abstract search algorithm, and so it is not amenable to its analytical method. Further research includes analytically proving the runtime of the algorithm, and exploring whether other quantum walk algorithms can be sped up using lackadaisical quantum walks.


\begin{acknowledgements}
	Thanks to Scott Aaronson for useful discussions. This work was supported by the U.S.~Department of Defense Vannevar Bush Faculty Fellowship of Scott Aaronson.
\end{acknowledgements}


\bibliographystyle{qinp}
\bibliography{refs}

\end{document}